\documentclass[prb,twocolumn,floatfix,showpacs,superscriptaddress]{revtex4}
\usepackage{graphics}
\usepackage{dcolumn}
\usepackage{bm}
\usepackage{epsfig}
\usepackage{color}
\usepackage{amsmath,amssymb}
\newcommand{\abs}[1]{\lvert #1\rvert}
\newcommand{\mean}[1]{\langle#1\rangle}
\newcommand{\ur}[1]{\,\mathrm{#1}}
\begin{document}
\title{Pressure-induced two-step spin crossover in double-layered elastic model}
\author{Daisuke Taniguchi}
\email{taniguchi@astron.s.u-tokyo.ac.jp}
\affiliation{Department of Astronomy, The University of Tokyo, 7-3-1 Hongo, Bunkyo-ku, Tokyo 113-0033, Japan}
\author{Jun Okabayashi}
\affiliation{Research Center for Spectrochemistry, The University of Tokyo, 7-3-1 Hongo, Bunkyo-ku, Tokyo 113-0033, Japan}
\author{Chisa Hotta}
\affiliation{Department of Basic Sciences, The University of Tokyo, 3-8-1 Komaba, Meguro, Tokyo 153-8902, Japan}
\date{\today}
\begin{abstract}
We study the two-step spin crossover in a double-layered elastic model based on transition metal complexes each taking high spin (HS) and low spin (LS) states. 
Here, only the simplest elastic interactions between adjacent molecules are considered 
and the system is exposed to the external pressure within the framework of $NPT$-Monte Carlo method. 
As a certain amount of pressure is applied, 
the first order thermal transition between uniform HS and LS phases transforms to a two-step transition with an emergent intermediate spin (IS) phase, 
where the HS and LS molecules are paired face to face between layers and form diagonally striped clusters within the layer. 
The difference in the size of HS and LS molecules is reflected both in the elastic interactions and in the enthalpy, 
and the IS phase could gain the latter over the loss of the former by significantly reducing its volume.
The present pressure effect is interpreted to the chemical one in double-layered transition metal materials, 
which actually reveals a variety of multistep spin crossover transitions relevant to our numerical result. 
\end{abstract}
\maketitle
\narrowtext
\section{Introduction}
Over the years, there has been a growing demand to utilize the spin-crossover (SCO) materials as molecular devices, 
such as ultrafast switches, reversible nanoscale memories, and sensors of temperature and pressure~\cite{review1,review2}.  
One of the advantages toward the device applications is the numbers of existing SCO compounds available with a variety of active working ranges, 
as the switching between high-spin (HS) and low-spin (LS) states can be easily controlled by temperature variation, 
pressure~\cite{p1,p2,p3}, light irradiation~\cite{opt1,opt2}, or magnetic field~\cite{h1,h2,h3,h4}. 
The SCO complexes consist of molecular magnets containing transition metal (TM) ions surrounded by the octahedral ligands, 
and the manipulation of the ligand field on TMs varies the degrees of splitting of energy level, which drives the HS to LS and vice versa 
on a single molecular unit. 
A conventional simplified model maps the HS and LS states of the $i$-th molecule to a pseudo spin degrees of freedom, $s_{i}=\pm 1$, 
whose energy levels differ by $\Delta e=(D_0-k_BT \ln g)$, with $D_0$ a constant and $k_{B}T\ln g$ characterizes the entropy that 
stabilizes the HS state at high temperature. 
In the primitive picture, these pseudo spins form a noninteracting massive ensemble, 
and the competition between the energy and entropy terms controls $\Delta e$, and yields the crossover from a HS at high temperature to the LS at low temperature. 
Experimentally, the SCO does not remain a simple crossover but show a variety of transitions including the first order ones~\cite{sorai}. 
To explain such cooperative nature of the transition, 
fictitious Ising type interactions are introduced in a series of phenomenological studies, 
represented by the Wajnflasz-Pick (WP) model~\cite{wajnflasz1970etude, wajnflasz1971transitions, boukheddaden2000dynamical, boukheddaden2000one, miyashita2005structures}. 
Although these models have reproduced the overall qualitative features of the SCO, 
such as temperature hysteresis, the microscopic origin of the exchange interactions remains unclear. 

A realistic approach as an alternative was to focus on difference in the size of the HS and LS molecules by a few percent~\cite{Konig}, 
and to translate the stress caused by the local lattice distortion of the irregularly packed molecules to the elastic interactions~\cite{bari1972low,zimmermann1977model,nishino2007simple}. 
A more precise analysis showed that the local elastic stress due to the larger HS molecules propagates within the crystalline lattice 
and gradually drives the switching of the LS to HS phase~\cite{enachescu2010cluster}. 
Such local stress is also sensitive to the external pressure, and the pressure dependence of the SCO transition is also well explained in the elastic model~\cite{stoleriu2008elastic,rotaru2008size,konishi2008monte}.
A large volume change between HS and LS state typically by a few percent is important from the experimental side, 
and indeed, the actuators due to the huge spontaneous strain accompanied by the spin-state switching was proposed very recently~\cite{ncomm}. 

In the present paper, we focus on the effect of pressure on the material volume, with in mind both the chemical and external (physical) pressure, and show that the molecules of different sizes coexist in the single phase by clustering and shrinking its volume with the aid of pressure, which could be the cause of the two-step SCO transition. 
The coexistent phase is called the intermediate spin (IS) phase as it appears between the HS and LS phases.
The existence of distinct IS phase has been reported in several SCO materials~\cite{bonnet,bonnet2,molnar,sciortino,clements,kosose,murphy,sugaya}. 
These compounds are made of either di-iron (binuclear) or mono-iron systems, 
and in one of the former materials, Fe$^{\rm II}$(ethyl nicotinate)$_2$[Au$^{\rm I}$(CN)$_2$]$_2$, 
the pairing of bilayers may cause the strong aurophilic interactions, 
which is considered to be the origin of the recently found two-step or multi-step SCO~\cite{sugaya}. 
Such structural feature is modeled as a pressure effect on the double-layered system, and our results show that 
the pressure is actually indispensable to understand the two-step transition driven by the material volume. 

Previously, the simplest realization of IS phase in theories was to form a two sublattice structure of HS and LS. 
The accumulated studies on the classical spin models tell us that in the extended WP model, 
adding the ``ferromagnetic'' intra-sublattice interactions besides the ``antiferromagnetic'' inter-sublattice ones 
stabilizes such phase~\cite{bousseksou1992ising,boukheddaden2003dynamical,nishino2003arrhenius}. 
There are also cases where the Ising interactions are extended to the geometrically 
frustrated ones~\cite{boukheddaden2007theoretical,watanabe2016ordering} which are the analogues of the historically well-known ANNNI models~\cite{ANNNI}, 
generating a axially striped or disordered spin states consisting of the mixture of up and down spins~\cite{diep}. 
Some extensions of the elastic model to explain the two-step SCO were also proposed, 
e.g., the atom-phonon coupling model~\cite{nasser2004two}, the elastic model hybridized with Ising interaction~\cite{nishino2013effect}. 
However, again, all these studies remain phenomenological as there is no legitimate microscopic ground to include such direct or complex interactions. 

A more straightforward extension of the elastic model is given in Ref.~\onlinecite{paez2016elastic}; 
similar to the context of the WP model, they found that the next nearest neighbor elastic interactions could be a driving force of the appearance of the two sublattice IS phase.
Originally, the next nearest neighbor interaction in the elastic model was considered to play only a secondary role in order to keep the square lattice structure stable, 
and was excluded in the study based on the hexagonal or triangular lattices~\cite{enachescu2010cluster}. 
Whereas, in Ref.~\onlinecite{paez2016elastic}, 
the equilibrium relative positions of the molecules (under the elastic potential) are set to a certain range to prefer the formation of the square shaped HS sublattice which serves as a cage to accommodate the LS molecule. 
Only in such setup, the relatively strong next nearest neighbor interactions favor IS phase (which we confirmed in our calculation). 

In the present study, we get rid of any such assumptions and go back to one of the simplest elastic model by Konishi {\it et al.}~\cite{konishi2008monte}, 
but instead consider the double-layered systems and apply pressure, with in mind the recent observation of multi-step SCO in the double-layered materials. 
It turns out that the double-layered system behaves much more sensitive to the external pressure, 
optimizing its structure and volume, which could be one of the possible origins of the IS phase. 
The paper is organized as follows; 
in $\S$ 2, we explain the details of the model and method we developed in including the larger pressure effect than before. 
The results are shown in $\S$ 3, and the mechanism of the emergence of IS phase is discussed in $\S$ 4, in relevance to the experiment. 

\begin{figure}[tbp]
\centering 
\includegraphics[width=\columnwidth ]{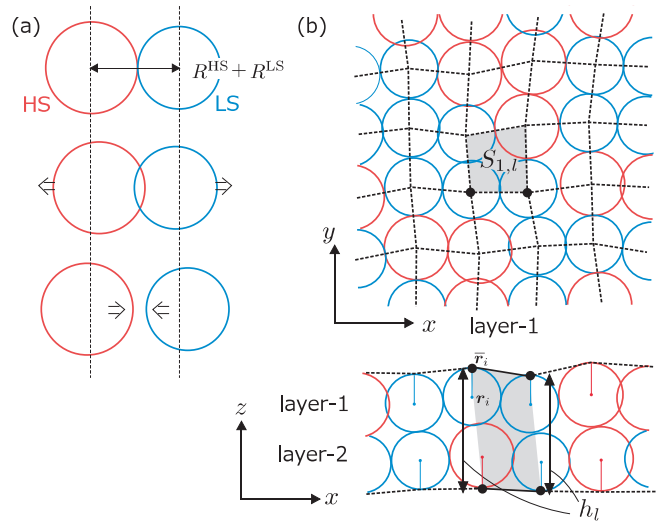}
\caption{(color online) (a) Schematic illustration of the effect of elastic interactions. 
The top panel shows the example when the adjacent HS and LS molecules are in the equilibrium distance, $R^{\mathrm{HS}}+R^{\mathrm{LS}}$. 
The second and third panels show the elastic force due to the displacement of molecules, which gives the elastic energy. 
(b) Schematic explanation of how we evaluate the volume in unit of a hexahedron. 
The upper panel shows a top view of the $xy$-plane, and the shaded square represents the $l$-th hexahedron of area $S_{1,l}$. 
The bottom panel is the cross section of the lattice in the $zx$-plane, where the height of the 
$l$-th hexahedron (shaded region) is given by the mean values of the length of these arrows.}
\label{fig:model}
\end{figure}
\section{Model and Method}
We adopt the elastic model which basically follows that of Konishi, {\it et al.}~\cite{konishi2008monte}, 
and consider two layers each accommodating $L\times L$ molecules. 
These molecules take either HS or LS, which is represented as spheres of large or small radius, 
$R^{\mathrm{HS}}$ or $R^{\mathrm{LS}}$, respectively. 
Their positions, $\bm r_i$, 
are smoothly varied while keeping an approximate square lattice structure within each layer. 
The Hamiltonian is given as, 
\begin{align}
\mathcal{H}&=\mathcal{H}_{0}+\mathcal{H}_{\mathrm{nn}}+\mathcal{H}_{\mathrm{nnn}} \\
&\mathcal{H}_{0}=\left(\frac{D_{0}}{2}-\frac{T}{2}\ln g\right)\sum _{i}s_{i} \\
&\mathcal{H}_{\mathrm{nn}}=\frac{k_{1}}{2}\sum_{\langle i,j\rangle }\left[\abs{\bm r_{i}-\bm r_{j}}-(R_{i}+R_{j})\right]^{2} \\
&\mathcal{H}_{\mathrm{nnn}}=\frac{k_{2}}{2}\sum_{\langle \langle i,j\rangle \rangle }\left[\abs{\bm r_{i}-\bm r_{j}}-\sqrt{2}(R_{i}+R_{j})\right]^{2}\text{.}
\end{align}
where $s_i=1$ and $-1$ represent the high and low spin states of the $i$-th molecule, respectively, 
and the indices run over $i=1$ to $N= 2L^2$. 
The on-site (single molecule) term, $\mathcal{H}_{0}$, consists of two terms; 
the first term represents the energy difference between the HS and LS levels, $D_0$, 
and the second term is the entropy difference that arises from the ratio of the degree of degeneracy, $g$, of the HS state against that of the LS state, 
which is introduced throughout the previous theoretical studies~\cite{wajnflasz1971transitions}. 
The competition of the two terms as a function of temperature, $T$, (setting the Boltzmann constant $k_B=1$) 
qualitatively reproduces the manipulation of the ligand-field splitting in TM, which is the overall origin of SCO; 
$\Delta e/2$, the coefficient of $s_i$ in $\mathcal{H}_{0}$, changes its sign at some temperature, 
which causes the switching between the HS state at high $T$ and LS state at low $T$. 
\par
In the rest of the terms, we take account of the elastic interaction between the molecules. 
Behind these interactions there is a harmonic oscillatory potential 
which takes a minimum when the distance between the neighboring two molecules, $i$ and $j$, becomes 
the summation of the radii of their spin states, $R^{\mathrm{HS}}$ or $R^{\mathrm{LS}}$, as shown in Fig.~\ref{fig:model}(a). 
As in Ref.~\onlinecite{konishi2008monte}, we consider the interactions on
bond connecting the nearest neighbor sites, $\langle i,j\rangle $, and the next nearest neighbor sites, 
$\langle \langle i,j\rangle \rangle $, 
while keeping the elastic constant of the latter much smaller than that of the former as, $k_1\gg k_2$. 
The elastic interactions between molecules of different layers are included in the same manner. 

We treat the model classically since we are dealing with the one- or two-step SCO materials which basically does not reveal any quantum effects.
While allowing $\bm r_i$ to take continuous values, 
the periodic boundary condition of $\bm r_i$ is imposed in the in-plane $xy$ directions. 
We perform the $NPT$-Monte Carlo (MC) method~\cite{mcdonald1969monte} for the isothermal-isobaric ensemble usually adopted to fluids, where $NPT$ represents the particle number, pressure, and temperature. 
In this method, the enthalpy $W=\mathcal{H}+PV$ is used instead of the energy (besides the correction term) 
in the original Metropolis algorithm in order to deal with the pressure effect. 
The details of our formulation is the same as in Ref.~\onlinecite{konishi2008monte}, besides two points; 
we include $g$ explicitly as described above, and determine $V$ precisely as explained below. 

In the usual $NPT$-MC method, 
the system length $l$ that gives the volume $V=l^3$ is taken as a typical length scale of the system, which is treated as 
MC parameters together with the normalized locations of molecules, $\bm r_i/l$. 
This treatment is valid in cases where the molecules are loosely packed, namely the pressure is small enough as in Ref.~\onlinecite{konishi2008monte}, 
which assumes that $l$ only slightly changes to optimize $\mathcal{H}+Pl^3$ during the MC processes. 
However, when the pressure is high enough and the molecules are tightly packed, varying $l$ and keeping $\bm r_i/l$ unchanged during the simulation 
would change the set of $\bm r_i$ significantly, which will vary both the elastic term and $PV$. 
Whereas, varying $\bm r_i/l$ and keeping $l$ will change only the elastic term. 
Thus, the two parameters are not independently tuning the two energy terms, but rather focusing on the adjustment of the elastic term, 
which hinders the proper optimization of the simulation. 
In other words, one could keep the location of $\bm r_i$ unchanged, while decreasing $l$ (and accordingly, increasing $\bm r_i/l$ simultaneously), 
so that $l^3$ does not reflect the proper volume of the system. 
To overcome this issue, we determine the volume more precisely based on the set of $\bm r_i$ according to the following steps 
(see also Fig.~\ref{fig:model}(b)); 
(i)~A set of positions of molecules, $\bm r_i$, forms an approximately square shaped lattice in both layers. 
Shifting these square shaped lattices along the $z$-direction to the surface of the layer, we define a set of $\bar{\bm r}_i$, as shown in the bottom panel of Fig.~\ref{fig:model}(b). Thus by pairing the closest face to face square units of $\bar{\bm r}_i$'s in the two layers, hexahedrons are formed. 
The system will be decomposed into $l=1 \sim L^2$ fragments of hexahedrons. 
(ii)~The area of the two faces (squares) of the $l$-th hexahedron, $S_{1,l}$ and $S_{2,l}$ belonging to the first and second layers, respectively, 
are measured by projecting the coordinates $\bar{\bm r}_i$ onto the $xy$-plane. 
(iii)~The height of the $l$-th hexahedron, $h_l$, is given as the mean length of the four edges connecting the two faces, after projecting them onto the $z$-axis. 
(iv)~Finally, the volumes of all the hexahedrons are summed up as, $V = \sum_l h_l(S_{1,l}+S_{2,l})/2$. 
Alternatively, one can precisely determine the volume of all the  hexahedrons. 
However, in the present double-layered system, the pressure is assumed to be imposed along the $z$ axis so that the above treatment 
could give a better evaluation of $PV$, which is the amount of work along the $z$ axis. 

The simulation is carried out at several fixed values of $P$, and the system is gradually cooled down from $k_{B}T=1.2$ to 0.0 ($0.4$ at higher pressure) in steps of order 0.01, 
which we regard as a single run of a cooling process, and vice versa in the heating process. 
Choosing the initial state as HS/LS state at high/low temperature for each run, we minimize 
the enthalpy by determining the distances between nearest neighbor sites in advance, and after that repeat the Monte Carlo process.
Once the system reaches the equilibrium at a fixed temperature, we measure the physical quantities by averaging over 10$^6$ Monte Carlo steps (MCSs). 
Then, we vary the temperature and start from the previous equilibrium state and repeat the process. 
At each temperature, $5\times 10^6$ MCS are discarded during the relaxation processes. 
We performed approximately 200--300 independent runs in the cooling process and 60 runs in the heating process at most. 
In several runs of the cooling process, the system is trapped to the IS state even at lowest temperature, where the structure is highly distorted. 
As such structure could be hardly relaxed except by annealing or releasing the pressure, we discarded 
the runs which have a HS fraction at $T=0.4$ higher than 0.01. 
%
\begin{figure}[tbp]
\centering 
\includegraphics[width=\columnwidth ]{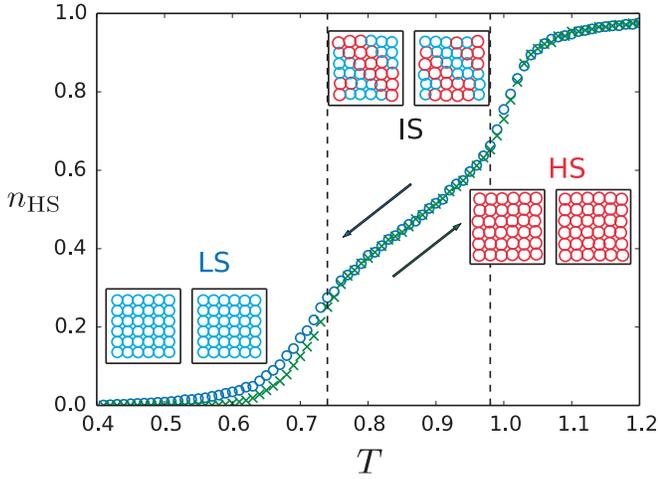}
\caption{(color online) Temperature dependence of HS fraction, $n_{\mathrm{HS}}^{}$, at $P=3$ and $N=72$. 
Blue circles and green crosses represent the cooling and heating processes, respectively. 
Inset panels show examples of the spin arrangements of upper and lower layers 
at $T=1.1$ (HS phase), $T=0.9$ (IS phase), and $T=0.5$ (LS phase).}
\label{fig:HSfraction}
\end{figure}
%
\begin{figure}[tbp]
\centering 
\includegraphics[width=0.75\columnwidth ]{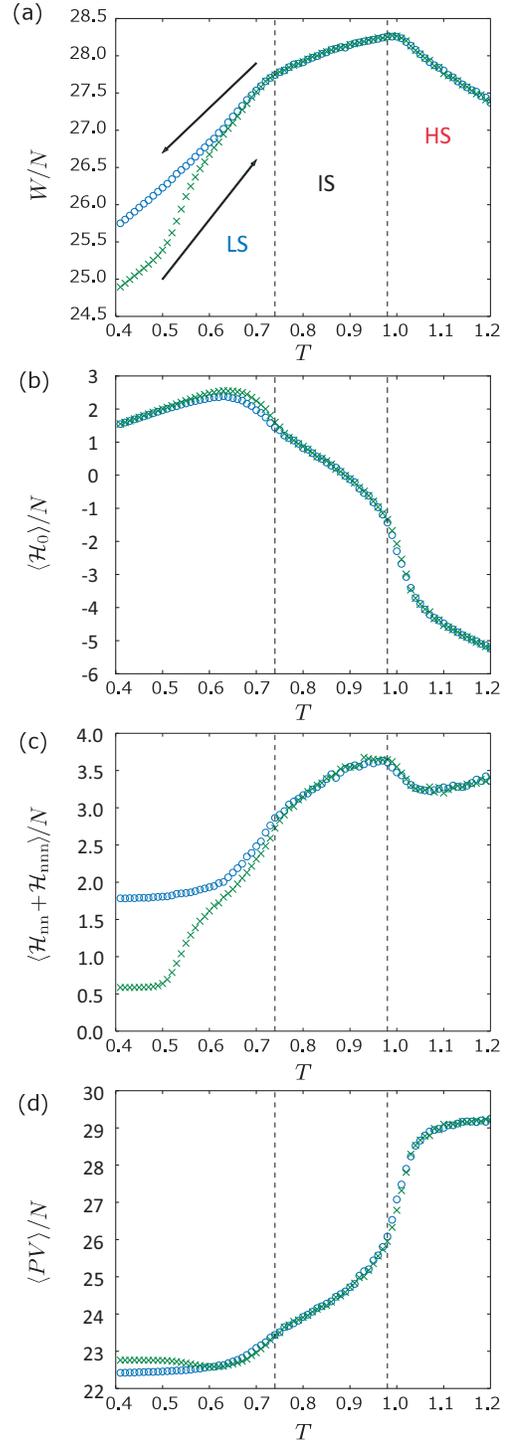}
\caption{(color online) Temperature dependence of (a) the enthalpy, $W$, 
(b) $\mean{\mathcal{H}_{0}}$ (contribution to free energy from a noninteracting part of the Hamiltonian), 
(c) the elastic energy, $\mean{\mathcal{H}_{\mathrm{nn}}+\mathcal{H}_{\mathrm{nnn}}}$, 
and (d) the $\mean{PV}$ term, where $W=\mean{\mathcal{H}_{0}}+\mean{\mathcal{H}_{\mathrm{nn}}+\mathcal{H}_{\mathrm{nnn}}}+P\langle V\rangle$. 
The cooling and heating processes written in blue circles and green crosses follow that of Fig.~\protect\ref{fig:HSfraction}. 
}
\label{fig:energy}
\end{figure}
%
\begin{figure}[tbp]
\centering 
\includegraphics[width=\columnwidth ]{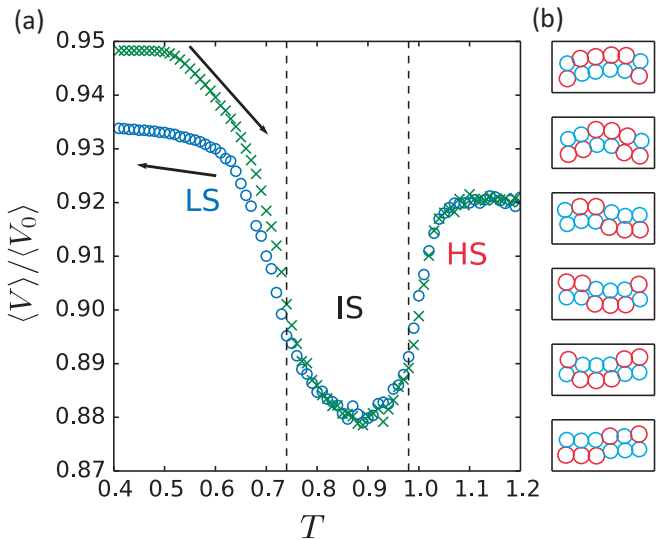}
\caption{(color online) (a) Calculated mean values of the system volume $\langle V\rangle$, against the normalized volume $\langle V_0\rangle$, 
which is the sum of the independent molecular volumes. 
(b) Molecular arrangement of two layers in the side view (plotting $L=6$ rows one by one) at $T=0.9$, 
where red and blue circles represent the HS and LS molecules, respectively.}
\label{fig:volumeshrink}
\end{figure}
%
\begin{figure}[tbp]
\centering 
\includegraphics[width=\columnwidth ]{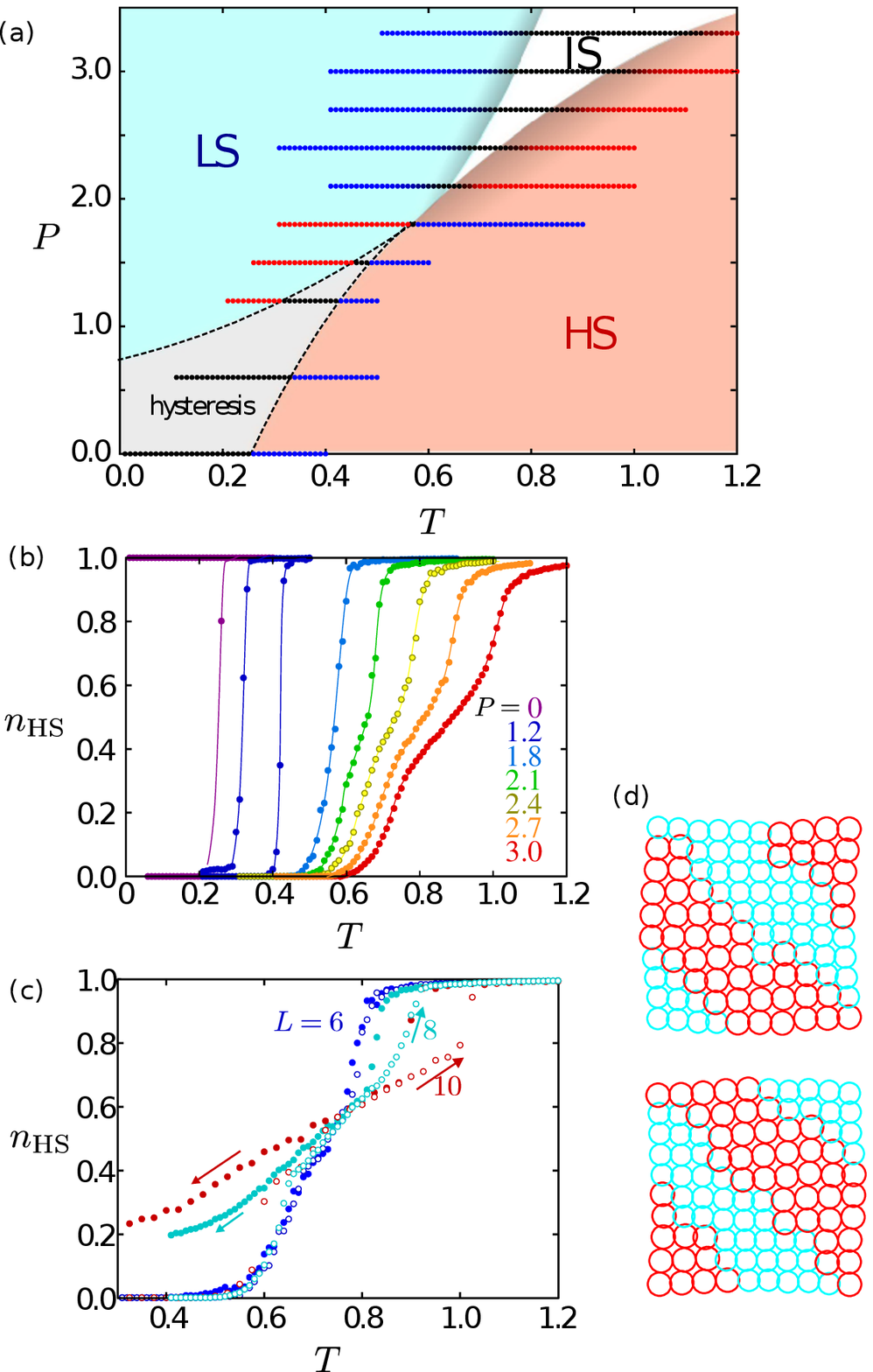}
\caption{(color online) (a) Phase diagram of a double-layered elastic model on the plane of pressure $P$ and temperature $T$, 
calculated for $N=72$, $k_1=10k_2=200$, $D_0=1$ and $\ln g=10$. 
(b) HS fraction, $n_{\mathrm{HS}}$, corresponding to the data points in panel (a). 
(c) $n_{\mathrm{HS}}^{}$ in the cooling (solid circle) and heating (open circle) processes at different system size, $L=6,8$ and 10. 
In the cooing process, the system is trapped to the metastable IS-like phase down to lowest temperature at larger system size. 
(d) Snapshots of the molecular arrangement in the IS phase obtained at $L=10$. 
}
\label{fig:phase}
\end{figure}
%
%
\section{Result}
We set the model parameters to $D_{0}=1$, $\ln g=10$, $R^{\mathrm{HS}}=1.1$, $R^{\mathrm{LS}}=1.0$, $k_{1}=200$, $k_{1}/k_{2}=10$, 
and impose the pressure up to $P=4$. 
In the previous work by Konishi, {\it et al.}~\cite{konishi2008monte} on the three dimensional cubic-like lattice, 
the parameters are chosen as $D_0=1$, $g=20$ (or $\ln g\sim 3$), and $k_1=10k_2\le 50$, 
in which case a HS--LS crossover of a transition takes place at $T\sim 0.3\text{--}0.8$ at $P\sim 0.01\text{--}0.5$, 
whereas for larger $k_1$, the transition disappears in the cooling process and the high spin phase remains down to $T=0$. 
We set $k_j$'s to four times stronger values in order to stabilize the two dimensional lattice structure. 
As the primary energy scale of the lattice is determined by $k_1$, the pressure needed to moderately influence the molecular arrangement 
is required to be four times larger than the previous studies, which we set to be $P=0\text{--}4$. 

The main results of the calculations are given on the $N=2\times 6^2=72$ molecules ($L=6$), which is relatively small compared to the previous studies. 
This is because it is difficult to retain a proper two dimensional layered structure under high pressure, as the coordinates could be varied freely even along the $z$-axis. For example, at $L=20$, the relaxation toward the proper structure could be easily hindered and the lattice structure collapses. 
We assume that the real material systems consist of stacking of our double-layered units. 
If we take account of such structural three-dimensionality by dealing with weakly coupled double layers, the above mentioned strucural instability shall be resolved, even when the effective pressure is imposed.

Figure \ref{fig:HSfraction} shows the HS fraction, $n_{\mathrm{HS}}^{}=(\mean{s_{i}}+1)/2$, in the cooling and heating process at $P=3$. 
One can identify the distinct IS phase at $T\sim 0.74\text{--}0.98$. 
The actual configuration of molecules in the IS structure reveals a diagonal stripe geometry of inter-layer HS-LS pairs of molecules. 
The crossover temperature, $k_BT_{cr}$ is roughly estimated by the contribution from the $\mathcal{H}_{0}$ and $PV$ terms; 
when taking account only of $\mathcal{H}_{0}$ it shall be scaled as $T_{cr} \sim D_0/\ln g$, 
at which the inversion of the relative location of the HS and LS energy levels occurs. 
The $PV$ term favors the LS state with smaller radius (smaller volume), so that it cooperates with $D_0$. 
Thus, we expect the correction to be included as $T_{cr} \sim (D_{0}+8P((R^{\mathrm{HS}})^{3}-(R^{\mathrm{LS}})^{3}))/\ln g$, which yields the value of $0.89$ consistent with our simulation. 

In order to understand the mechanism of transition between HS, IS, and LS, we measured the temperature dependence of enthalpy, 
$W=\mean{\mathcal{H}_{0}}+\mean{\mathcal{H}_{\mathrm{nn}}+\mathcal{H}_{\mathrm{nnn}}}+\mean{PV}$, as shown in Fig.~\ref{fig:energy}(a). 
It shows a sudden change at around the two transition points. 
This could be recognized as the crossings of three different enthalpy lines of different slopes belonging to the HS, LS, and IS phases. 
Let us separate the contributions of free energy, elastic potential, and $PV$ to $W$; 
In Fig.~\ref{fig:energy}(b), the contribution from $\mean{\mathcal{H}_{0}}$ to $W$ is shown, 
which represents the free energy of a single molecule on an average, and thus simply reflects the HS fraction. 
Notice that this term does not include the contribution of the entropy from the many body effect, which comes from the variation of configuration of molecules. 
The characteristic feature of the transition is visible in Figs. \ref{fig:energy}(c) and \ref{fig:energy}(d), 
which are the elastic potential energy, $\mean{\mathcal{H}_{\mathrm{nn}}+\mathcal{H}_{\mathrm{nnn}}}$, and $PV$ term, respectively. 
One finds that the IS phase has a loss in the elastic potential, but instead gains $PV$. 

The above results indicate that the volume $V$ shrinks in the IS state even by sacrificing the loss of the elastic potential, 
and leads to the relatively smaller $PV$, compared to HS and LS states. 
In order to visualize this tendency, we plot in Fig.~\ref{fig:volumeshrink}(a) the volume against the sum of the molecular volume, $\langle V \rangle /\langle V_0\rangle$, 
where $V_0=\sum_{j=1}^N (2R_i)^3$. The large dip at IS phase indicates that the molecules are particularly tightly packed by making use 
of the mixture of LS and HS molecules of different radius. 
 
The pressure dependence of the stability of the IS phase is examined in the phase diagram in Fig.~\ref{fig:phase}(a) on the plane of $P$ and $T$. 
The boundary of the IS is determined by the change in the slope of $n_{\mathrm{HS}}^{}$ shown in Fig.~\ref{fig:phase}(b). 
At lower $P$, the temperature ranges where IS appears becomes narrow, and finally, the IS phase disappears at around $P\sim 2$. 
Below that pressure, the system goes to the region where the direct transition between HS and LS states 
could be observed, accompanied by the large temperature hysteresis region, 
consistent with the results by Konishi, {\it et al.}~\cite{konishi2008monte}. 
At higher pressure, $P\gtrsim 3.6\text{--}4$, the square-like lattice structure is no longer maintained. 
 
Indeed, the higher the pressure the more sensitively the calculation depends on the initial condition, and the system could be easily trapped by 
a local minima with highly distorted (unphysical) molecular arrangements, particularly at the larger system size.
In order to check the size dependence of the results, we examined 
the temperature dependent HS fraction, $n_{\mathrm{HS}}^{}$ at $P=2.4$, through the cooling and heating processes, 
for $L=6,$ $8$ and $10$ as given in Fig.~\ref{fig:phase}(c). 
At $L=8$ and 10, during the cooling process, the system stays in the IS phase and does not transform to the LS phase even 
at temperature lower than $T \lesssim 0.5$. This indicates that the IS phase remains as a metastable state down to low temperature. 
Once the system is trapped to this metastable state, it is rather difficult to rearrange the system by a local flipping and moving of spins in the MC calculation. 
In fact, for cases where the IS state is absent, namely when the system undergoes a direct first order transition from the HS to LS phase, 
the similar behavior is observed~\cite{konishi2008monte}; the system sustains a HS state when $k_1$ is large, namely the height of the elastic potential well becomes deep. 
At larger system size, the number of metastable state increases, 
so that it is much difficult in the actual calculations to transform from IS to LS in the cooling process. 
Figure \ref{fig:phase}(d) shows the snapshot configuration of the IS phase at $L=10$. 
The diagonal stripe structure is present, and one may find a number of choices of the HS and LS configurations 
that may slightly change the ratio of HS and LS at around $n_{\mathrm{HS}}^{}\sim 0.5$, that may contribute to the entropy of the system, 
which may be the reason for the stable IS phase at larger $L$.

\section{Discussion}
\label{ssec:origin}
\begin{figure}[tbp]
\centering 
\includegraphics[width=\columnwidth ]{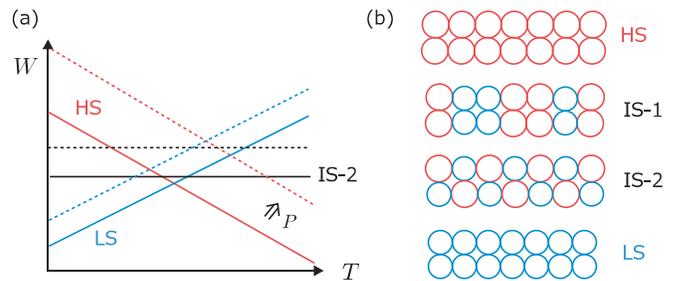}
\caption{(color online) (a) Schematic illustration of simplified description of $W^{\mathrm{st}}$ in Eq.~(\ref{enthalpy}). 
The broken and solid lines represent the ones at ambient and finite pressures, respectively. 
(b) The idealized alignment of molecules in the HS, IS, and LS phases, where 
the IS-2 has smaller volume than the IS-1. 
As the elastic energy is larger in the IS-2 state, the gain in the $PV$ term due to smaller $V$ 
is considered to be responsible for the downshift of $W^{\mathrm{IS\text{-}2}}$ from $W^{\mathrm{IS\text{-}1}}$. 
}
\label{fig:enthalpy}
\end{figure}

Let us discuss the origin of the competition between the $PV$ and the elastic term in more detail by roughly estimating the enthalpy by hand, and comparing it with the numerical results. 
For simplicity, the elastic potential, $\mean{\mathcal{H}_{\mathrm{nn}}+\mathcal{H}_{\mathrm{nnn}}}$, 
and the volume of the three phases are assumed to be almost constant at the given fixed configuration in each phase which we denote here as, 
$E_{k}^{\mathrm{st}}$ and $V^{\mathrm{st}}$, for $\mathrm{st}=$ HS, IS, and LS, respectively. 
The free energies of the (noninteracting) molecules are denoted as $E_{0}^{\mathrm{st}}\equiv \mean{\mathcal{H}_{0}}$. 
While the actual fraction of HS molecules of the IS phase vary at around $n_{\mathrm{HS}}^{}\sim 0.4\text{--}0.6$, we consider the case of $n_{\mathrm{HS}}^{}=0.5$, namely, 
half of the molecules have LS state and the rest remains as the HS state. 
These approximations give the enthalpy, 
\begin{align}
W^{\mathrm{st}}&\sim E_{0}^{\mathrm{st}} +E_{k}^{\mathrm{st}}+PV^{\mathrm{st}},
\label{enthalpy} \\
&E_{0}^{\mathrm{HS/LS}}=\pm \left(\frac{D_{0}}{2}-\frac{T}{2}\ln g\right)N \\ 
&E_{0}^{\mathrm{IS}}=0,
\end{align}
which are shown schematically in Fig.~\ref{fig:enthalpy}(a). Their functional form roughly reproduces our numerical results in Fig.~\ref{fig:energy}(a). 
The relationship, $W^{\mathrm{IS}}<W^{\mathrm{HS}}$ and $W^{\mathrm{IS}}<W^{\mathrm{LS}}$, is realized 
when the pressure and temperature satisfy the following two conditions;
\begin{align}
&P>\frac{2E_{k}^{\mathrm{IS}}-E_{k}^{\mathrm{HS}}-E_{k}^{\mathrm{LS}}}{V^{\mathrm{HS}}+V^{\mathrm{LS}}-2V^{\mathrm{IS}}}\text{,} 
\label{cond1}
\\
&-(V^{\mathrm{LS}}\!-\!V^{\mathrm{IS}})P+\frac{D_{0}N}{2}+(E_{k}^{\mathrm{IS}}\!-\!E_{k}^{\mathrm{LS}})<\frac{N\ln g}{2}T \nonumber \\
&\qquad < (V^{\mathrm{HS}}-V^{\mathrm{IS}})P+\frac{D_{0}N}{2}-(E_{k}^{\mathrm{IS}}-E_{k}^{\mathrm{HS}}) .
\label{cond2}
\end{align}
Both conditions can be fulfilled when $V^{\mathrm{IS}}$ is significantly small enough compared to $V^{\mathrm{HS}}$ and $V^{\mathrm{LS}}$. 


Suppose that each molecule occupies the volume of a cube, $\mean{(2R_{i})^{3}}$, by considering that the neighboring molecules do not overlap, whose configuration is given in IS-1 in Fig.~\ref{fig:enthalpy}(b).
Then the relation, $V^{\mathrm{HS}}+V^{\mathrm{LS}}-2V^{\mathrm{IS-1}}=0$, holds and Eq.~(\ref{cond2}) is no longer fulfilled, 
thus IS becomes unstable due to the large $E_{k}^{\mathrm{IS}}$.
In our double layer, the volume of the IS phase could be further suppressed by considering that the molecules in the two layers placed face to face 
always form pairs of HS and LS, and the HS and LS molecules are aligned in the staggered manner in each plane as in the previous studies 
(IS-2 type of configuration in Fig.~\ref{fig:enthalpy}(b)). 
The approximate volume of IS-2 shrinks to $N(R^{\mathrm{HS}}+R^{\mathrm{LS}})^{3}$ and the condition $V^{\mathrm{HS}}+V^{\mathrm{LS}}-2V^{\mathrm{IS}}>0$ is satisfied. 
However, the gain in the $PV$ term is still subtle so that it is not enough to always satisfy Eq.~(\ref{cond2}) 
by compensating for the energetic disadvantage of the IS phase in $E_{k}^{\mathrm{st}}$. 
In our numerical results, the arrangement of HS and LS molecules shown in Fig.~\ref{fig:volumeshrink}(b) 
is realized as a result of balance between the pressure and the elastic interactions. 
In such case, even though some of the bonds become only slightly shorter than $\mean{R_{i}+R_{j}}$, 
the total volume shrinks by warping the surface so as to minimize the airspace between molecules. 
Indeed, as one can see in Fig.~4(a), 
the evaluated $\langle V_{\mathrm{st}}\rangle$ of $\mathrm{st}=$HS, LS, and IS, against $V_0=\sum_i (2R_{i})^{3}$ is significantly small 
in the IS phase, supporting our estimation. 

We remark that we found no evidence of an existing IS phase in single-layered two-dimensional system within our model on a $6\times 6$ lattice. 
In fact, if we do not take account of the volume suppression characteristic of the double layer, there is no reason to form stripe patterns; 
it is more favorable to have disordered HS and LS spin arrangement, as it has many spatial patterns that contribute to the large (many body) 
entropy gain of order-$N$. 
However, such disordered patterns of HS and LS has large elastic energy loss so that it is also difficult to overwhelm the uniform HS and LS phases. 

Finally, let us discuss the relevance to the actual bilayer SCO materials, 
Fe$^{\rm II}$(ethyl nicotinate)$_2$[Au$^{\rm I}$(CN)$_2$]$_2$~\cite{sugaya} and Fe(pyridine)$_2$ [Ag(CN)$_2$]$_2$~\cite{rodrigues}. 
In these materials, the octahedral ligand based on Au and Ag ions seem to play important role in the emergence of the two-step or multi-step SCO transition. 
The  Au-Au distances in the bilayers, $\sim 3.1\,\text{\AA, }$ is smaller by $15\%$ from the sum of the van der Waals radii of Au ($3.60\,\text{\AA }$), 
indicating that the tightly packed crystal structures possibly due to the strong Au-Au interactions work as an effective chemical pressure. 
In the organic materials, the chemical pressure is often interpreted to the real external pressure 
by comparing the experimentally realized phases. 
In Fe(3-methylpyridine)$_2$[Ni(CN)$_4$], the two-step transition appears when the pressure of order $100\ur{MPa}$ is applied~\cite{molnar}, 
thus the chemical pressure of the above mentioned compounds may also amount to that order. 
In Ref.~\onlinecite{slimani2013}, 
the model parameters are determined as $D_0=900\ur{K}$ by comparing the difference of the molar enthalpy between HS and LS state, 
$\Delta W= D_0 N_A$, with the typical experimental value of $5\text{--}20\ur{kJ/mol}$, where $N_A$ is the Avogadro constant. 
In our case, by replacing $D_0$ with $D_0+8P((R^{\mathrm{HS}})^3-(R^{\mathrm{LS}})^3)$ and by setting $R^{\mathrm{HS}}=1.1\ur{nm}$, $R^{\mathrm{LS}}=1\ur{nm}$, 
we obtain, $D_0=250\ur{K}$, the value $P=1$ corresponding to $3.4\ur{MPa}$, and  $T=1$ to $250\ur{K}$, accordingly. 
Whereas, if we simply adopt $\Delta W= D_0 N_A$, the pressure $P=1$ becomes $12\ur{MPa}$. 
We mention that the value of $P$ required to stabilize the IS phase increases if we set $k_1$ to larger values. 
Also, if the inter-layer potentials is taken account of in our model to keep the layered structure more stable, 
one could examine larger values of $P$. 
Experimentally, a larger temperature hysteresis is found in the IS--LS transition compared to the HS--IS one~\cite{sugaya}. 
This is also the case with our results in Fig.~\ref{fig:HSfraction}, while the quantitative comparison is still out of reach as they depend 
on the model parameters. 

\section{Summary}
We considered the double-layered system, whose layer consists of molecules forming an approximate square lattice structure 
with in mind the transition metal (TM) compounds showing two-step spin crossover (SCO) transitions. 
The high and low spin states (HS and LS) of a TM ion are described as the up and down pseudo-spin states, 
and the switching between the two in a molecular unit is basically controlled by the temperature through the parameter, 
$\Delta e=(D_0-T\ln g)$, which is the energy difference 
between the two levels that converts its sign when varying the temperature. 
To take account of the cooperative nature of the SCO transition, namely the first order HS to LS transitions and the two-step transition which is of our focus, 
the interactions between molecular states are included as elastic interactions linear to the displacement from the equilibrium distance between the molecules. 
This model already succeeded in realizing the pressure induced first order transition between spatially uniform HS phase to LS phases~\cite{konishi2008monte}.
A two-step transition with an intermediate spin (IS) phase of a checkerboard (two sublattice) pattern of HS and LS molecules was observed~\cite{paez2016elastic} 
by tuning the lattice equilibrium position of the molecules and by taking account of the relatively larger elastic 
intra-sublattice (next nearest neighbor) interactions. 
As the combination of HS and LS molecules basically increases the elastic energy, such state is stabilized by a rather fine tuning of the model. 
We actually confirmed in the present calculation that without such constraint on the equilibrium position, the checkerboard IS is not stable 
even by increasing the next nearest neighbor interactions. 

In reality, the system volume shrinks at the HS to LS phase transition typically by 5$\%$, 
and the energetics that reflect this large volume change may play an important role in the transition. 
Also, in some of the TM materials exhibiting two-step transition, the sheet layers of the material show a warping 
due to chemical decoration from ligand molecules, which means that the molecules are much more tightly packed~\cite{sugaya}.
Whereas, in the previous analysis on the elastic model, the molecules are loosely packed; e.g., in the checkerboard type IS phase of Ref.~\onlinecite{paez2016elastic}, 
the LS molecules are embedded in the airspace of a HS sublattice in its nearly equilibrium position. 
In our model, instead of tuning the model parameters to favor such particular IS phase, we set our parameters to cases where 
only the first order transitions between uniform HS and LS phases takes place at low pressure. 
Here, the volume of the system is precisely determined in order to properly reflect the subtle differences in the local molecular arrangement to the $PV$ term. 
Then, by increasing the pressure we find a particular IS phase in which the HS and LS molecules gather in diagonally striped patterns within the layer, 
while forming pair of HS and LS between layers. 
This phase is stabilized by the gain in $PV$ by optimizing the arrangement of HS and LS, while sacrificing the loss of the elastic potential energy to some extent. 
To be more precise, the energy scale of $PV$ need to be larger than the elastic potential, and also $PV$ and elastic potential to be larger than $D_0$ to have the IS phase.

\begin{acknowledgements}
This work is supported by Grant-in-Aid for Scientific Research (Nos. 17909321, 17924266, 17895051) from the Ministry of Education, Science, Sports and Culture of Japan. 
\end{acknowledgements}

\label{totalpage}
\end{document}